\documentclass{article}

\usepackage{arxiv}

\usepackage{amsmath,amssymb}
\usepackage{graphicx}
\usepackage{wrapfig}
\usepackage{array}
\usepackage[algoruled,linesnumbered,algo2e,vlined]{algorithm2e}
\usepackage{amssymb}
\usepackage{url}
\usepackage{multirow}
\usepackage{cite}

\usepackage{algorithm}
\usepackage{algorithmic}

\usepackage{comment}

 \newtheorem{definition}{Definition}
 \newtheorem{theorem}{Theorem}

 \newenvironment{proof}{\trivlist \item[\hskip \labelsep{\bf proof.\/}]}{\hspace{\fill}$\Box$\endtrivlist }

\title{Privacy-Preserving Feature Selection with Fully Homomorphic Encryption}

\author{
Shinji Ono
\And
Jun Takata 
\And
Masaharu Kataoka 
\And
Tomohiro  I 
\And
Kilho Shin$^*$ 
\And
Hiroshi Sakamoto\\ \\
Kyushu Institute of Technology, 680-4 Kawazu, Iizuka, Fukuoka 820-8502, Japan \\
\texttt{\{ono.shinji514, takata.jun903, kataoka.masaharu403\}@mail.kyutech.jp}\\
  \texttt{ \{tomohiro, hiroshi\}@ai.kyutech.ac.jp} \\ \\
$^*$Gakushuin University, 1-5-1 Mejiro, Toshimaku, Tokyo 171-8588, Japan \\ 
\texttt{kilhoshin314@gmail.com}
}

\begin{document}
\maketitle
\begin{abstract}
For the feature selection problem, we propose an efficient privacy-preserving algorithm.
Let $D$, $F$, and $C$ be data, feature, and class sets, respectively, where the feature value $x(F_i)$ and 
the class label $x(C)$ are given for each $x\in D$ and $F_i \in F$.
For a triple $(D,F,C)$, the feature selection problem is to find a consistent and minimal subset $F' \subseteq F$,
where `consistent' means that, for any $x,y\in D$, $x(C)=y(C)$ if $x(F_i)=y(F_i)$ for $F_i\in F'$,
and `minimal' means that any proper subset of $F'$ is no longer consistent. 
On distributed datasets, we consider feature selection as a privacy-preserving problem:
Assume that semi-honest parties $\textsf A$ and $\textsf B$ have their own personal $D_{\textsf A}$ and $D_{\textsf B}$. 
The goal is to solve the feature selection problem for $D_{\textsf A}\cup D_{\textsf B}$ without revealing their privacy.
In this paper, we propose a secure and efficient algorithm based on fully homomorphic encryption, 
and we implement our algorithm to show its effectiveness for various practical data.
The proposed algorithm is the first one that can directly simulate the CWC (Combination of Weakest Components) 
algorithm on ciphertext, which is one of the best performers for the feature selection problem on the plaintext.
\end{abstract}

\section{Introduction}

\subsection{Motivation}
Feature selection is one of the most common problems in machine learning.
For example, the human genome contains of 3.1 billion base pairs, only a few dozens of which are thought to affect a specific disease. 
Various machine learning algorithms make use of favorable features extracted from such sparse data.

Consider a data set $D$ associated with a feature set $F$ and a class variable $C$, where
all feature values $x(F_i)$ $(F_i\in F)$ and the corresponding class label $x(C)$ are defined for each data $x\in D$.
In Table~\ref{fig1}, for example, we show a concrete example.
Given a triple $(D,F,C)$, 
the feature selection problem is to find a minimal $F'\subseteq F$ that is relevant to the class $C$.
The relevance of $F'$ is evaluated, for example, by $I(F';C)$, which measures the mutual information between $F'$ and $C$.
On the other hand, $F'$ is minimal, if any proper subset of $F'$ is no longer consistent.

\begin{table*}[t]
\begin{center}
\caption{
An example dataset shown in~\cite{Shin2017}.
} 
\label{fig1}
\begin{tabular}{c|ccccc|c}
\hline
$D$ &$F_1$ & $F_2$ & $F_3$ &$F_4$ & $F_5$ & \;$C$\; \\ \hline
$x_1$& 1 & 0 & 1 & 1 &  1 & 0 \\
$x_2$& 1 & 1 &0 &0 &0& 0 \\
$x_3$& 0 &0& 0& 1& 1& 0 \\
$x_4$& 1& 0 &1& 0& 0& 0 \\
$x_5$& 1& 1&1& 1 &0& 1 \\
$x_6$& 0 &1 &0& 1& 0 &1 \\
$x_7$& 0& 1& 0& 0 &1 &1 \\
$x_8$& 0 &0 &0& 0& 1& 1 \\ \hline
\;$I(F_i;C)$\;& \;0.189\; & \;0.189\;& \;0.049\;& \;0.000\;& \;0.000\; &  \\ 
\hline
\end{tabular}
\end{center}
\end{table*}

To the best of our knowledge, the most common method for identifying favorable features is to choose
features that show higher relevance in some statistical measure.
Individual feature relevance can be estimated using statistical measures such as mutual information and Bayesian risk.
For example, at the bottom row of Table~\ref{fig1}, the mutual information score $I(F_1;C)$ of each feature $F_i$ to class labels is described.
We can see that $F_1$ is more important than $F_5$, because $I(F_1;C)>I(F_5;C)$.
$F_1$ and $F_2$ of Table~\ref{fig1}  will be chosen to explain $C$ based on the mutual information score.
However, a closer examination of $D$ reveals that $F_1$ and $F_2$ cannot uniquely determine $C$.
In fact, we find $x_2$ and $x_5$ with $x_2(F_1)=x_5(F_1)$ and $x_2(F_2)=x_5(F_2)$ but $x_2(C)\neq x_5(C)$.
On the other hand, we can see that $F_4$ and $F_5$ uniquely determine $C$ using the formula 
$C=F_4\oplus F_5$ while $I(F_4;C)= I(F_5;C) =0$.
As a result, the traditional method based on individual features relevance scores misses the right answer.

So we concentrate on the concept of consistency: $F'\subseteq F$ is called to be consistent if, for any $x,y\in D$, $x(F_i)=y(F_i)$ for all $F_i\in F'$ implies $x(C)=y(C)$.
In machine learning research, consistency-based feature selection has received a lot of attention~\cite{Shin2011,Shin2009,Zhao2007,Liu1998,Almuallim1994}.
CWC (Combination of Weakest Components)~\cite{Shin2009} is the simplest of such consistency-based feature selection algorithms, and even though CWC uses the most rigorous measure, 
it shows one of the best performances in terms of accuracy as well as computational speed compared to other methods~\cite{Shin2017}.

To design a secure protocol for feature selection,
we focus on the framework of homomorphic encryption.
Given a public key encryption scheme $E$, let $E[m]$ denote a ciphertext of integer $m$;
if $E[m+n]$ can be computed from $E[m]$ and $E[n]$ without decrypting them,
then $E$ is said to be {\em additive} homomorphic, and if $E[mn]$ can also be computed, then $E$ is said to be {\em fully} homomorphic.
Furthermore, modern public-key encryption must be {\em probabilistic}: when the same message $m$ is encrypted multiple times, 
the encryption algorithm produces different ciphertexts of $E[m]$.

Various homomorphic encryption schemes have been proposed to satisfy those homomorphic properties over the last two decades.
The first additive homomorphic encryption was proposed by Paillier~\cite{Paillier1999}. 
{\em Somewhat} homomorphic encryption that allows a sufficient number of additions and a limited number of multiplications
has also been proposed~\cite{Boneh2005,Brakerski2012,Attrapadung2018}, 
and we can use these cryptosystems to compute more difficult problems, such as the inner product of two vectors.
Gentry~\cite{Gentry2009} proposed the first fully homomorphic encryption (FHE) with unlimited number of additions and multiplications, 
and since then, useful libraries for fully homomorphic encryption have been developed, particularly for  
bitwise operations and floating-point operations.
TFHE~\cite{Chillotti2020,TFHE} is known as a fastest fully homomorphic encryption that is optimized for bitwise operations. 

For the private feature selection problem, we use TFHE to design and implement our algorithm. 
In this case, we assume two {\em semi-honest} parties $\textsf A$ and $\textsf B$: each party complies the protocol but tries to 
infer as much as possible about the secret from the information obtained.
The parties have their own private data $D_{\textsf A}$ and $D_{\textsf B}$ and they jointly compute advantageous features 
for $D_{\textsf A}\cup D_{\textsf B}$ while maintaining their privacy.
The goal is to jointly compute the CWC algorithm result on $D=D_{\textsf A} \cup D_{\textsf B}$ 
without revealing any other information.

It should be a realistic requirement,
if one wants to draw some conclusions from data 
that are privately distributed over more than one parties.   
Multi-party computation (MPC) can provide effective technical solutions to realize this requirement in many cases.  
In MPC, certain computation which essentially rely on the distributed data is performed 
through cooperation among the parties.
In particular, 
fully homomorphic encryption (FHE) is one of the critical tools of MPC.
One of the most significant advantages of FHE-based MPC is thought to be that
FHE realizes outsourced computation in a simple and straightforward manner:
Parties encrypt their private data with their public keys
and send the encrypted data to a single trusted party with sufficient computational power to perform the required computation;
Although the computational results by the trusted party may be wrong, 
if some malicious parties send wrong data,
honest parties are at least convinced that their private data have not been stolen
as far as the cryptosystem used is secure.  
In contrast,
when a party shares his/her secret with other parties to perform MPC, 
even if it uses a secure secret sharing scheme,
collusion of a sufficient number of compromised parties may reveal the party's secret.
In general,
it is hard to prove the security of MPC protocols
for the situation where we cannot deny the existence of active malicious parties,
and hence,
the security is very often proven assuming that all the parties are at worst semi-honest.
In the reality, however,
even this relaxed assumption is hard to hold.
Thus,
the property that a party can protect its private data
only relying on its own efforts should be counted as an important advantage of FHE-based MPC.

On the other hand,
the current implementations of FHE are thought to be significantly inefficient,
and consequently, their ranges of application is actually limited.
This is currently true,
but may not be true in the future:
The Goldwasser-Mmicali (GM) cryptosystem~\cite{Goldwasser1984} is thought as the first scheme with provable security;
Unfortunately,
because the GM cryptosystem encrypts data in a bit-wise manner,
it has turned out not to have sufficient efficiency in time and memory to be used in the real world;
In 2001, however, 
RSA-OAEP was finally proven to have both provable security and realistic efficiency~\cite{Fujisaki2001,Bellare1994},
and is widely used through SSL/TLS. 
Thus,
studying FHE-based MPC does not merely have theoretical meaning, 
but also will yield significant contributions in terms of application to the real world in the future.

In this paper,
we propose a MPC protocol which relies on FHE-based outsourced computation as well as
mutual cooperation among parties.
The target of our protocol is to perform the computation of CWC,
a feature selection algorithm known to be accurate and efficient,
preserving privacy of participating parties.  
If we fully perform CWC by FHE-based outsourced computation,
we have to pay unnecessarily large costs in time in the phase of sorting features of CWC.
Therefore,
in our proposed scheme,
we add ingenuity so that two parties cooperate with each other to sort features efficiently.

Converting CWC into its privacy-preserving version based on different primitives of MPC,
for example based on secret sharing techniques, is not only interesting
but also useful both in theory and in practice.
We will pursue this direction as well as our future work.

\subsection{Our contribution and related work}

\begin{table*}[t]
\begin{center}
\caption{
Time and space complexities of the baseline and improved algorithms for secure CWC,
where $k$ is the number of features and $m,n$ are the numbers of positive and negative data, respectively.
We assume that the time of respective operation (e.g., encryption/addition/multiplication/comparison) in FHE is $O(1)$.
} 
\label{result}
\begin{tabular}{ccc}
\hline
Algorithm & Time & Space \\ \hline
CWC on plaintext~\cite{Shin2009} & $O(kmn+k\log k)$ & $O(kmn)$ \\
secure CWC (baseline) & $O(kmn\log k+k\log^2 k)$ & $O(kmn)$ \\
improved  & $O(kmn+k\log ^2k+k\log k\log mn)$ & $O(kmn)$\; \\
\hline
\end{tabular}
\end{center}
\end{table*}

Table~\ref{result} summarizes the complexities of proposed algorithms in comparison to the original CWC on plaintext.
The {\em baseline} is a naive algorithm that can simulate the original CWC~\cite{Shin2009} over ciphertext using TFHE operations. 
The bottleneck of private feature selection exists in the sorting task over ciphertext, as we mention in the related work below.
Our main contribution is the improved algorithm, shown as `improved',  which significantly reduces the time complexity caused by the sorting task.
We also implement the improved algorithm and demonstrate its efficiency through experiments in comparison to the baseline.

In this section, we discuss related work on the private feature selection as well as the benefits of our method.
Rao~\cite{Rao2019} et al. proposed a homomorphic encryption-based private feature selection algorithm.
Their protocol allows the additive homomorphic property only, which invariably leaks statistical information about the data.
Anaraki and Samet~\cite{Anarakia2020} proposed a different method based on the rough set theory, but their method 
suffers from the same limitations as Rao et al., and neither method has been implemented.
Banerjee et al.~\cite{Banerjee2011}, and Sheikhalishahi and Martinellil~\cite{Sheikhalishahi2017} have proposed 
MPC-based algorithms that guarantee security by decomposing the plaintext into shares
as a different approach to the private feature selection, while achieving cooperative computation. 
Li et al.~\cite{Li2021} improved the MPC protocol on aforementioned flaw and demonstrated its effectiveness through experiments. 

These methods avoid partially decoding under the assumption that the mean of feature values provides a good criterion for feature selection.
This assumption, however, is heavily dependent on data. 
The most important task in general feature selection is feature-value-based sorting, 
and CWC and its variants~\cite{Shin2009,Shin2011,Shin2017} demonstrated the effectiveness of sorting 
with the consistency measure and its superiority over other methods. 
On ciphertext, this study realizes the sorting-based feature selection algorithm (e.g., CWC).

We focus on the learning decision tree by MPC~\cite{Abspoel2021} as another study that employs sorting for private machine learning, 
where the sorting is limited to the comparison of $N$ values of fixed-length in $O(N\log^2N)$ time by a sorting network. 
In the case of CWC, however, the algorithm must sort $N$ data points, each of which has a variable-length of up to $M$, 
so a naive method requires $O(MN\log N + N\log^2N)$ time. 
Our algorithm reduces this complexity to $O(MN+N\log^2N+N\log N\log M)$, that is significantly smaller than the naive algorithm depending on $M$ and $N$.
By experiments, we confirm this for various data including real datasets for machine learning.


\section{Preliminaries}

\subsection{CWC algorithm over plaintext}
We generally assume that the dataset $D$ associated with $F$ and $C$ contains no errors, i.e.,
if $x(F_i)=y(F_i)$ for all $i$, $x(C)=y(C)$.
When $D$ contains such errors, they are removed beforehand and $D$ contains not more than
one $x\in D$ with the same feature values.

In Algorithm~\ref{algo1}, we describe the original algorithm for finding a minimal consistent features. 
Given $D$ with $F_i$ and $C=\{0,1\}$, 
a data $x\in D$ of $x(C)=1$ is referred to as a positive data and $y\in D$ of $y(C)=0$ is referred to as a negative data.
Let $n$ represent the number of positive data and $m=|D|-n$.
Let $x_p$ represent the $p$-th positive data $(1\leq p\leq n)$ and $y_q$ represent the $q$-th negative data $(1\leq q\leq m)$.
Then, the bit string $\mathbf{B}_i$ of length $nm$ is defined by:
$\mathbf{B}_i[m(p-1)+q] = 0$ if $x_p(F_i)=y_q(F_i)$ and $\mathbf{B}_i[m(p-1)+q] = 1$ otherwise.
$\mathbf{B}_i[m(p-1)+q] = 0$ means that $F_i$ is not consistent with the pair $(x_p,y_q)$ 
because $x_p(F_i)=y_q(F_i)$ despite $x_p(C)\neq y_q(C)$.
Recall that $F_i$ is said to be consistent only if $x(F_i)=y(F_i)$ implies $x(C)=y(C)$ for any $x,y\in D$.
As a result, $||\mathbf{B}_i||$ is defined to be the number of $1$s in $\mathbf{B}_i$.

For a subset $F'\subseteq F$,
$F'$ is said to be consistent, if for any $p\in[1,n]$ and $q\in [1,m]$,
there exists $i$ such that $F_i\in F'$ and $\mathbf{B}_i[m(p-1)+q]=1$ hold.
CWC uses this to remove irrelevant features from $F$ in order to build a minimal consistent feature set\footnote{
Finding a smallest consistent feature set is clearly NP-hard due to an obvious reduction from the minimum set cover.
}.

\begin{algorithm}   
\caption{The algorithm CWC for plaintext}                         
\label{algo1}                          
\begin{algorithmic}[1] 
\STATE Input: A dataset $D$ associated with features $F=\{F_1,\ldots,F_k\}$ and class $C=\{0,1\}$.
\STATE Output: A minimal consistent subset $S\subseteq F$.
\STATE Sort $F_1,\ldots,F_k$ in the incremental order of $||\mathbf{B}_i||$.
\STATE Let $\pi$ be the sorted indices of $\{1,\ldots,k\}$.
\FOR{$i=1,\ldots,k$} 
\IF{$F\setminus \{F_{\pi[i]}\}$\text{ is consistent}}
\STATE update $F\leftarrow F\setminus \{F_{\pi[i]}\}$
\ENDIF
\ENDFOR
\end{algorithmic}
\end{algorithm}

Table~\ref{fig2} shows an example of $D$, and Table~\ref{fig_bs} shows the corresponding $\mathbf{B}_i$.
Consider the behavior of CWC in this case.
All $\mathbf{B}_i$ $(1\leq i\leq 4)$ are computed as preprocessing.
Then, the features are sorted by the order 
$||\mathbf{B}_2||=5\leq ||\mathbf{B}_4||=5\leq ||\mathbf{B}_3||=6\leq ||\mathbf{B}_1||=8$ and $\pi=(2,4,3,1)$.
By the consistency order $\pi$, CWC checks whether $F_{\pi[i]}$ can be removed from the current $F$.
Using the consistency measure, CWC removes $F_2$ and $F_4$ and the resulting $\{F_1,F_3\}$ is the output.
In fact, we can predict the class of $x$ by the logical operation $\overline{x(F_1)}\wedge x(F_3)$.

\begin{table*}[t]
\begin{center}
\caption{
An example dataset $D$ with $F=\{F_1,F_2,F_3,F_4\}$ and $C=\{0,1\}$.
The data consists of two positive data $\{ x_1, x_2\}$ and
five negative data $\{ y_1, y_2, y_3, y_4, y_5 \}$.
} 
\label{fig2}

  \begin{minipage}[t]{.45\textwidth}
    \begin{center}
      \begin{tabular}[t]{c|cccc|c} \hline
        \;$x_i\in D$\; &\;$F_1$\; & \;$F_2$\; & \;$F_3$\; & \;$F_4$\; & \;$C$\; \\ \hline
        $x_1$& 0 & 1 & 1 & 0 & $1$\\
        $x_2$& 0& 0&1& 1 &$1$ \\ \hline
      \end{tabular}
    \end{center}
  \end{minipage}
  \hfill
  \begin{minipage}[t]{.45\textwidth}
    \begin{center}
      \begin{tabular}[t]{c|cccc|c} \hline
        \;$y_i\in D$\; &\;$F_1$\; & \;$F_2$\; & \;$F_3$\; & \;$F_4$\; & \;$C$\; \\ \hline
        $y_1$& 1 & 0 &1 &0 &0 \\
        $y_2$& 1 &1& 0& 0& 0 \\
        $y_3$& 0& 1 &0& 1& 0 \\
        $y_4$& 1 &0 &1& 0& 0 \\
        $y_5$& 1& 1& 0& 0 &0 \\ \hline
      \end{tabular}
    \end{center}
  \end{minipage}
\end{center}
\end{table*}

\begin{table*}[t]
\begin{center}
\caption{
The bit string $\mathbf{B}_i$ for the example dataset $D$ of Table~\ref{fig2}.
Each column $(x_p, y_q)$ is $0$ iff $x_p(F_i) = y_q(F_i)$.
For example, $\mathbf{B}_1 = (1,1,0,1,1,1,1,0,1,1)$ because 
$x_p(F_1) = y_q(F_1)$ only for the two pairs $(x_1, y_3)$ and $(x_2, y_3)$.
} 
\label{fig_bs}
\begin{tabular}{c|cccccccccc}
\hline
$\mathbf{B}_i$ & {\scriptsize $(x_1, y_1)$ } & {\scriptsize $(x_1, y_2)$ } &{\scriptsize $(x_1, y_3)$ } & {\scriptsize $(x_1, y_4)$} &{\scriptsize  $(x_1, y_5)$}  & {\scriptsize $(x_2, y_1)$} & {\scriptsize $(x_2, y_2)$ } & {\scriptsize $(x_2, y_3)$ } & {\scriptsize $(x_2, y_4)$ } & {\scriptsize  $(x_2, y_5)$}  \\ \hline
$\mathbf{B}_1$ & 1 & 1 & 0 & 1 & 1 & 1 & 1 & 0 & 1 & 1\\
$\mathbf{B}_2$ & 1 & 0 & 0 & 1 & 0 & 0 & 1 & 1 & 0 & 1\\
$\mathbf{B}_3$ & 0 & 1 & 1 & 0 & 1 & 0 & 1 & 1 & 0 & 1\\
$\mathbf{B}_4$ & 0 & 0 & 1 & 0 & 0 & 1 & 1 & 0 & 1 & 1\\
\hline
\end{tabular}
\end{center}
\vspace{-2mm}
\end{table*}

\subsection{Security model}

\subsubsection{Indistinguishable random variables}
Let $\mathbb{N}$ denote the set of natural numbers.
A function $\epsilon:\mathbb{N}\to[0,1]$ is called {\em negligible},
if $\forall c>0,\; \exists k,\; \forall n\geq k,\; \epsilon(n)<1/n^c$.
Let $X=\{X_n \mid k\in \mathbb N\}$ and $Y=\{Y_k\mid k\in\mathbb N\}$ be
sequences of random variables 
such that $X_k$ and $Y_k$ are defined over the same sample space.
We say that $X$ and $Y$ are \emph{indistinguishable}, denoted by $X\equiv_c Y$, 
if, and only if,
$\Pr[X_n = Y_n]$ is a negligible function.  

\subsubsection{Security of multi-party computation (MPC)}

Although the discussion of this section can be extended to MPC schemes which involve more than two parties,
just for simplicity,
we focus on the case where only two parties are involved.

A two-party protocol is a pair $\Pi = ({\cal P}_1,{\cal P}_2)$ of PPT Turing machines with input and random tapes.
Let $x_i$ be an input of ${\cal P}_i$ and $y_i$ be an output of ${\cal P}_i$, respectively.

We assume a {\em semi-honest adversary} ${\cal A}$ and consider a protocol $({\cal A},{\cal P}_2)$, replacing ${\cal P}_1$ in $\Pi$ by ${\cal A}$,
where $\cal A$ takes $x_1$ as input and apparently follows the protocol.
Let ${\tt REAL}_{\Pi,\cal A}(x_1,x_2)$ denote the random variable representing the output $(y_1,y_2)$ of $({\cal A},{\cal P}_2)$,
and we define the class ${\tt REAL}_{\Pi,\cal A} = \{{\tt REAL}_{\Pi,\cal A}(x_1,x_2)\}_{x_1,x_2} = \{y_1,y_2\}_{x_1,x_2}$.

On the other hand, let $\cal F$ denote the functionality that the protocol $\Pi$ is trying to realize, i.e.,
$\cal F$ is a PPT that simulates the honest $({\cal P}_1,{\cal P}_2)$ so that
${\cal F}(x_1,x_2)\equiv ({\cal P}_1(x_1),{\cal P}_2(x_2))$.
Here, we assume a completely reliable third party, denoted by ${\cal F}$.
In this {\em ideal} world, for this $\cal F$ and any adversary $\cal B$ acting as ${\cal P}_1$
with input $x'_1$ possibly $x'_1\neq x_1$, we define the random variable
${\tt IDEAL}_{{\cal F}, \cal B}(x_1,x_2) = ({\cal B}(x_1,{\cal F}_1(x'_1,x_2),{\cal F}_2(x'_1,x_2)))$, where
${\cal F}_i(\cdot,\cdot)$ denotes the $i$-th component of the output of ${\cal F}(\cdot,\cdot)$ for $i=1,2$.
Similarly, we denote the class ${\tt IDEAL}_{{\cal F}, {\cal B}} = \{{\tt IDEAL}_{{\cal F}, {\cal B}}(x_1,x_2)\}_{x_1,x_2}$.

Using such random variables, we define the security of protocol $\Pi$ as follows.

\begin{definition}\label{security}
It is said that a protocol $\Pi$ securely realizes a functionality ${\cal F}$ if
for any attacker $\cal A$ against $\Pi$, there exists an adversary $\cal B$,
${\tt REAL}_{\Pi,\cal A}\equiv_c {\tt IDEAL}_{{\cal F}, \cal B}$ holds.
\end{definition}

The definitions stated above can be intuitively explained as follows.
Exactly conforming to the protocol, a semi-honest adversary $\cal A$ plays the role of ${\cal P}_1$ to steal any secrets.
The information sources which $\cal A$ can take advantage of are the following three:
\begin{enumerate}
  \item The input tape to ${\cal P}_1$;
  \item the conversation with ${\cal P}_2$;
  \item the execution of the protocol.
\end{enumerate}
While the information that $\cal A$ can obtain from the first and third sources are exactly $x_1$ and $y_1$ respectively,
we call the information from the second source a \emph{view}.
To denote it, we use the symbol $\mathsf{View}_{{\cal P}_1}$.

Since the protocol enevitably requires that 
$\mathcal A$ obtains the information of $x_1$ and $y_1$,
the security of the protocol questions about
what $\cal A$ can obtain in addition to what can be computationally inferred from $x_1$ and $y_1$.
If there exists such information, its source must be $\mathsf{View}_{{\cal P}_1}$.  

The security criterion of \emph{simulatability} requires that
$\mathsf{View}_{P_1}$ can be simulated on input of $x_1$ and $y_1$.
To be formal,
there exists a PPT Turing machine \textsf{Sim}
that outputs a view on input of $x_1$ and $y_1$
such that the output view cannot be distinguished from 
$\mathsf{View}_{{\cal P}_1}$ by any PPT Turing machine.
When $\mathsf{View}_{{\cal P}_1}$ is simulatable,
we see that
\textsf{Sim} can generate by itself
what \textsf{Sim} can obtain from $\mathsf{View}_{P_1}$.
Therefore,
\textsf{Sim} cannot cannot obtain any information
in addition to what \textsf{Sim} can compute from $x_1$ and $y_1$. 

\subsubsection{IND-CPA}

\emph{Indistinguishability against chosen plaintext attack} (IND-CPA) is
an important criterion for secrecy of a public key cryptosystem.
We let $\Pi = ({\tt Gen},{\tt Enc},{\tt Dec})$ denote a public key cryptosystem
consisting of key generation, encryption, and decryption algorithms.
To describe IND-CPA,
we introduce the \emph{IND-CPA game} played between an adversary $\mathcal A$ and
an oracle $\mathcal O$:
$\mathcal A$ is a PPT Turing machine,
and $k$ is the security parameter.

\begin{enumerate}
\item $\mathcal O$ generates a public key pair $(sk,pk)\gets {\tt Gen}(1^k)$.
\item $\mathcal A$ generates two messages $(m_0,m_1)$ of the same length arbitrarily and
  throws a query $(m_0, m_1)$ to $\mathcal O$.
\item On receipt of $(m_0, m_1)$,
  $\mathcal O$ selects $b \in\{0, 1\}$ uniformly at random, 
  computes $c = {\tt Enc}(pk, m_b)$ and replies to $\mathcal A$ with $c$.
\item 
  $\mathcal A$ guesses on $b$ by examining $c$ and
  outputs the guess bit $b'$.
\end{enumerate}

We view $b$ and $b'$ as random variables whose underlying probability space is defined
to represent the choices of the public key pair, $b$ and $b'$.  
The \emph{advantage} of the adversary $\mathcal A$ is defined as follows
to represent the advantage of $\mathcal A$ over tossing a fair coin
to guess $\mathcal O$'s secret $b$:
\[
  \mathsf{Adv}_\mathcal A = 2 \cdot \Pr[b' = b] - 1.
\]
When we let
\[
  \Pr[b' = 0 \vert b = 0] = \frac 12 + \alpha_0\ \text{and}
  \Pr[b' = 1 \vert b = 1] = \frac 12 + \alpha_1, 
\]
we have
\[
  \mathsf{Adv}_\mathcal A = \alpha_0 + \alpha_1
\]
This definition of the advantage is consistent with the common definition found in many textbooks:
\[
  \mathsf{Adv}_\mathcal A =
  \Pr[b' = 0 \vert b = 0] - \Pr[b' = 1 \vert b = 0]
\]

\begin{definition}\label{CPA}
A public key cryptosystem $\Pi$ is secure in the sense of IND-CPA, or simply IND-CPA secure, 
if $\mathsf{Adv}_\mathcal A$ as a function in $k$ is a negligible function.
\end{definition}

\subsection{TFHE: a faster fully homomorphic encryption}

The proposed private feature selection is based on FHE.
We review the TFHE~\cite{TFHE}, one of the fastest libraries for bitwise addition (this means XOR `$\oplus$')
and bitwise multiplication (AND `$\cdot$') over ciphertext.
On TFHE, any integer is encrypted bitwise:
For $\ell$-bit integer $m=(m_1,\ldots,m_\ell)$, we denote its bitwise encryption by
$E[m] \equiv (E[m_1],\ldots,E[m_\ell])$, for short.
These bitwise operations are denoted by $f_\oplus (E[x], E[y]) \equiv E[x\oplus y]$ and $f_\cdot (E[x], E[y]) \equiv E[x\cdot y]$
for $x,y\in\{0,1\}$ and the ciphertexts $E[x]$ and $E[y]$.
The same symbol is used to represent an encrypted array.
For example, when $x$ and $y$ are integers of length $\ell$ and $\ell'$, respectively,
$E(x,y)$ denotes 
\[E[x,y]\equiv (E[x],E[y]) \equiv (   (E[x_1],\ldots,E[x_\ell]),  ( E[y_1],\ldots,E[y_{\ell'}])).\]

TFHE allows all arithmetic and logical operations via the elementary operations $E[x\oplus y]$ and $E[x\cdot y]$.
In this section, we will go over how to build the adder and comparison operations.
Let $x,y$ represent $\ell$-bit integers and $x_i,y_i$ represent the $i$-th bit of $x,y$ respectively.
Let $c_i$ represent the $i$-th carry-in bit and $s_i$ is the $i$-th bit of the sum $x+y$.
Then, we can get $E[x+y]$ by the bitwise operations of ciphertexts using 
$s_i = x_i\oplus y_i\oplus c_i$ and $c_{i+1}=(x_i\oplus c_i)\cdot (y_i\oplus c_i)\oplus c_i$.
We can construct other operations like subtraction, multiplication, and division based on the adder.
For example, $E[x-y]$ is obtained by $E[x+(-y)]$, where $(-y)$ is the bit complement of $y$ 
obtained by $y_i\oplus 1$ for all $i$-th bit.
On the other hand, we examine the comparison.
We want to get $E[x<?y]$ without decrypting $x$ and $y$ where $x<?y = 1$ if $x<y$ and $x<?y = 0$ otherwise.
We can get the logical bit for $x<?y$ as the most significant bit of $x+(-y)$ over ciphertexts here.
Similarly, for the equality test, we can compute the encrypted bit $E[x=?y]$.

Adopting those operations of TFHE, we design a secure multi-party CWC.
In this paper, we omit the details of TFHE (see e.g.,~\cite{Chillotti2020,TFHE}).

We should note that the secrecy of TFHE definitely impacts the security of our scheme.
In fact, in our two-party feature selection scheme,
the party \textsf B sends his/her inputs in an encrypted form to the party \textsf A,
and \textsf A performs the computation of feature selection on the encrypted inputs.
If the encrypted inputs could be easily cracked,
any ingenious devices to secure the scheme would be meaningless.

Therefore,
in designing our scheme,
it was a matter of course to require our FHE cryptosystem to be IND-CPA secure.
In fact,
TFHE is known to be IND-CPA secure.
Regarding this,
we should note the following
\begin{itemize}
\item By definition,
  encryption with an ID-CPA cryptosystem is probabilistic.
  That is,
  the result $E[x]$ of encryption unpredictably differs
  every time when the encryption is performed.
  For this reason, 
  by $E[x \vert t]$,
  we denote a ciphertext generated at time $t$.
  In particular,
  the notation of $E[x \vert\ast]$ means that the ciphertext has been generated
  at the time different from any other encryption events.
\item
When we consider the IND-CPA security of an FHE cryptosystem,
we should note that
the way how the oracle $\mathcal O$ generates $c$ with $D(c) = m_b$ is not unique.
For example,
the oracle may computes $c$ from two ciphertexts of additive shares of $m_b$, 
say $E[r]$ and $E[m_b \oplus r]$,  by $c = f_\oplus(E[r], E[m_b \oplus r])$.
The IND-CPA security of an FHE cryptosystem
should require that $\mathcal A$ cannot guess $b$ with effective advantage,
no matter how $c$ has been generated.
This, however, holds,
if the result of performing $E[x], f_\oplus(E[x], E[y])$ and $f_\cdot(E[x], E[y])$ distributes uniformly,
and TFHE is known to satisfy this condition.
\end{itemize}

\section{Algorithms}

\subsection{Baseline algorithm}

We present the baseline algorithm, a privacy-preserving variant of CWC.
In this subsection, we consider a two-party protocol, in which a party $\textsf A$ has his private data and 
outsources CWC computation to another party $\textsf B$,
but the baseline algorithm is easily extended to more than two data owners case, e.g.,
parties $\textsf A$ and $\textsf C$ send their private data to party $\textsf B$ using $\textsf A$'s public key.
During the computation, party $\textsf B$ should not gain other information than the number $n$ of positive data, 
the number $m$ of negative data and the number $k$ of features.
It should be noted that party $\textsf A$ can hide the actual number of data by inserting dummy data and 
telling $\textsf B$ the inflated numbers $n$ and $m$.
Dummy data can be distinguished by adding an extra bit that indicates the data is a dummy if the bit is $1$.
The values of features and dummy bits of data in each class are encrypted by $\textsf A$'s public key and sent to $\textsf B$.

The baseline algorithm consists of three tasks:
Computing encrypted bit string $E[\mathbf{B}_i]$,
sorting $E[\mathbf{B}_i]$'s and
executing feature selection on $E[\mathbf{B}_i]$'s.
In the baseline algorithm, all inputs are encrypted and they are not decrypted until the computation is completed.
Thus, for simplicity, we omit the notation $E$ in the following presentation.

\subsubsection{Computing $\mathbf{B}_i$}
We can compute $\mathbf{B}_i[m(p-1)+q]$ by $(x_p(F_i) \oplus y_q(F_i)) \vee x_p(d) \vee y_q(d)$, where
$x_p(d)$ and $y_q(d)$ represent the dummy bits for data $x_p$ and $y_q$, respectively.
$(x_p(F_i) \oplus y_q(F_i))$ becomes $0$ iff $F_i$ is inconsistent for the pair of $x_p$ and $y_q$.
Since we want to ignore the influence of dummy data, 
the part ``$\vee x_p(d) \vee y_q(d)$'' is added to make the whole value $1$ (meaning that it is consistent) when one of $x_p$ and $y_q$ is a dummy.
It takes $O(kmn)$ time and space in total.

\subsubsection{Sorting $\mathbf{B}$'s}
We can compute $\Vert\mathbf{B}_i\Vert$ in encrypted form by summing up values in $\mathbf{B}_{i}$ in $O(mn \log (mn))$ time
(noting that each operation on integers of $\log (mn)$ bits takes $O(\log (mn))$ time).
Instead, we can set an upper bound $b_{\mathit{max}}$ of the bits used to store consistency measure to reduce the time complexity to $O(mnb_{\mathit{max}})$.

Then, sorting $\mathbf{B}$'s in the incremental order of consistency measures can be accomplished
using any sorting network in which comparison and swap are performed in encrypted form 
without leaking information about feature ordering.
It should be noted that in this approach, the algorithm must spend $\Theta(mn + \log k)$ time to swap (or pretend to swap)
two-bit strings and original feature indices of $\log k$ bits regardless that two features are actually swapped or not.
Because this is the most complex part of our baseline algorithm, we will demonstrate how to improve it.
Using AKS sorting network~\cite{Ajtai1983} of size $O(k \log k)$, the total time for sorting $\mathbf{B}_i$'s is 
$O(mnb_{\mathit{max}} + (mn + b_{\mathit{max}} + \log k)k \log k)$.

In our experiments, we employ a more practical sorting network of Batcher's odd-even mergesort~\cite{Batcher1968} of size $O(k \log^2 k)$.
A a simple oblivious radix sort~\cite{Hamada2014} in $O(k\log k)$ algorithm under the assumption that
the bit length of each integer is constant was recently proposed.

\subsubsection{Selecting features}
Let $(F_{\pi(1)}, \ldots, F_{\pi(k)})$ be the sorted list of features.
We first compute a sequence of bit strings $(Z_2, \ldots, Z_k)$ of length $mn$ each such that
$Z_i[h] = \bigvee_{j = i+1}^{k} \mathbf{B}_{\pi(j)}[h]$ for any $2 \le i \le k$ and $1 \le h \le mn$,
namely $Z_i$ is the bit array storing cumulative or of each position $h$ for $\mathbf{B}_{\pi(i+1)}, \mathbf{B}_{\pi(i+2)}, \ldots, \mathbf{B}_{\pi(k)}$.
Note that $Z_i[h] = 0$ indicates that the set $\{ F_{\pi(i+1)}, F_{\pi(i+2)}, \ldots, F_{\pi(k)} \}$ of features is inconsistent w.r.t.\ 
a pair $(x_p, y_q)$ satisfying $h = m (p-1) + q$, and
$\{ F_{\pi(i+1)}, F_{\pi(i+2)}, \ldots, F_{\pi(k)} \}$ is inconsistent iff the bit string $Z_i$ contains $0$.
See Table~\ref{fig_Z} for $Z$'s in our running example.
The computation requires $O(kmn)$ time and space.

\begin{table*}[t]
\begin{center}
\caption{
Sorted $\mathbf{B}$'s for the example dataset $D$ of Table~\ref{fig2} and the corresponding $Z_i$'s.
} 
\label{fig_Z}
\scalebox{0.85}[0.85]{ 
\begin{tabular}{c|c|lcccccccccc|lcccccccccc}
\hline
$i$ & $\pi(i)$ & $\mathbf{B}_{\pi(i)}$ & & & & & & & & & & & $Z_i$ & & & & & & & & & & \\ \hline
$1$ & $2$      & $\mathbf{B}_2 = $ & 1 & 0 & 0 & 1 & 0 & 0 & 1 & 1 & 0 & 1 & $Z_1 = $ & 1 & 1 & 1 & 1 & 1 & 1 & 1 & 1 & 1 & 1\\
$2$ & $4$      & $\mathbf{B}_4 = $ & 0 & 0 & 1 & 0 & 0 & 1 & 1 & 0 & 1 & 1 & $Z_2 = $ & 1 & 1 & 1 & 1 & 1 & 1 & 1 & 1 & 1 & 1\\
$3$ & $3$      & $\mathbf{B}_3 = $ & 0 & 1 & 1 & 0 & 1 & 0 & 1 & 1 & 0 & 1 & $Z_3 = $ & 1 & 1 & 0 & 1 & 1 & 1 & 1 & 0 & 1 & 1\\
$4$ & $1$      & $\mathbf{B}_1 = $ & 1 & 1 & 0 & 1 & 1 & 1 & 1 & 0 & 1 & 1 & $Z_4 = $ & 0 & 0 & 0 & 0 & 0 & 0 & 0 & 0 & 0 & 0\\
\hline
\end{tabular}
}
\end{center}
\end{table*}

We simulate Algorithm~\ref{algo1} on encrypted $\mathbf{B}$'s and $Z$'s for feature selection.
Furthermore, we use two $0$-initialized bit arrays, $R$ of length $k$ and $S$ of length $mn$.
$R[i]$ is meant to store $1$ iff the $i$-th feature (in sorted order) is selected.
$S$ is used to keep track of the cumulative or for the bit strings of the currently selected features.
Namely, $S[h]$ is set to $\bigvee_{\alpha = 1}^{\ell} \mathbf{B}_{\pi(j_{\alpha})}[h]$ if $\ell$ features $\{ F_{\pi(j_{1})}, \ldots, F_{\pi(j_{\ell})} \}$ have been selected at the moment.

Assume that we are in the $i$-th iteration of the for loop of Algorithm~\ref{algo1}.
Note that, at the moment, $F$ contains features $\{ F_{\pi(i)}, F_{\pi(i+1)}, \ldots, F_{\pi(k)} \}$ and currently selected features,
and $F \setminus \{F_{\pi(i)}\}$ is consistent iff $\bigwedge_{h = 1}^{mn}(Z_{i}[h] \vee S[h])$ is $1$.
Because we keep $F_{\pi(i)}$ in $F$ iff $F \setminus \{F_{\pi(i)}\}$ is inconsistent,
the algorithm sets $R[i] = \neg \bigwedge_{h = 1}^{mn}(Z_{i}[h] \vee S[h])$.
After computing $R[i]$, we can correctly update $S$ by $S[h] \leftarrow S[h] \vee (R[i] \wedge \mathbf{B}_{\pi(i)}[h])$ for every $1 \le h \le mn$ in $O(mn)$ time.
Therefore, the total computational time is $O(kmn)$.

\subsubsection{Summing up analysis}

The sorting step takes $O(mnb_{\mathit{max}} + (mn + b_{\mathit{max}} + \log k)k \log k)$ time.
Because CWC works with any consistent measure, we do not need to use $\Vert\mathbf{B}_{i}\Vert$ in full accuracy, so 
we assume that $b_{\mathit{max}}$ is set to be a constant.
Under the assumption, we obtain the following theorem.

\begin{theorem}
For the two party feature selection problem,
we can securely simulate CWC in $O(kmn \log k + k \log^2 k)$ time
and $O(kmn)$ space without revealing the private data of the parties
under the assumption that TFHE is secure. 
\end{theorem}
\begin{proof}
According to the discussion above, computing $\mathbf{B}_i$ for all features takes $O(kmn)$ time and space,
sorting features takes $O(mnb_{\mathit{max}} + (mn + b_{\mathit{max}} + \log k)k \log k) = O(kmn \log k + k \log^2 k)$ time, and
selecting features takes $O(kmn)$ time.

Finally, party $\textsf B$ computes in $O(k \log k)$ time an integer array $P$ with $P[h] = R[h] \cdot \pi(h)$, 
which stores the original indices of selected features.
In outsourcing scenario, party $\textsf B$ simply sends $P$ to party $\textsf A$ as the result of CWC.
In joint computing scenario, party $\textsf B$ randomly shuffles $P$ to conceal $\pi$ to $\textsf A$.
As a result, we can securely simulate CWC in $O(kmn \log k + k \log^2 k)$ time and $O(kmn)$ space.
\end{proof}

\subsection{Improvement of secure CWC}

\begin{algorithm}   
\caption{Improved secure CWC between parties $\textsf A$ and $\textsf B$}                         
\label{new_sort}                          
\begin{algorithmic}[1] 
\STATE  {\bf Preprocessing:} \\
Party $\textsf A$ has $E_{\textsf B}[{\mathcal F}] =E_{\textsf B}[{\mathcal F}_1,\ldots,{\mathcal F}_k]$
for ${\mathcal F}_i = (F_i,\Vert\mathbf{B}_i\Vert,\mathbf{B}_i)$ encrypted with the party $\textsf B$'s public key,
where each data $x$ encrypted at time $0$ as $E_{\textsf B}[x \vert 0]$.
\STATE {\bf Party $\textsf A$:} \\
Generates $r_i$ for $i=1,\ldots,n$ uniformly at random.\\
Sends $(E_{\ \textsf B}[{\mathbf B}_i + r_i \vert 1],E_{\textsf A}[r_i\vert 1])$ for $i=1,\ldots,n$.
\STATE {\bf Party $\textsf A$:} \\
Calculates $E_{\textsf B}[i\vert 2]$ for $i=1,\ldots,n$. \\
Securely sorts $(E_{\textsf B}[F_i\vert 0],E_{\textsf B}[\Vert{\mathbf B}_i\Vert \vert 0],E_{\textsf B}[i\vert 2])$ for $i=1,\ldots,n$ 
in increasing order of $\Vert{\mathbf B}_i\Vert$.\\
As a result, obtains $(E_{\textsf B}[F_{i_j}\vert 3],E_{\textsf B}[\Vert{\mathbf B}_{i_j}\Vert \vert 3],E_{\textsf B}[{i_j}\vert 3])$ for $j=1,\ldots,n$.\\
Generates a permutation $\pi \in \mathfrak S_n$ uniformly at random and memorizes it. \\
Sends $(E_{\textsf B}[i_{\pi(1)}\vert 3], \ldots, E_{\textsf B}[i_{\pi(n)}\vert 3])$ .
\STATE {\bf Party $\textsf B$:} \\
Decrypts $(i_{\pi(1)},\ldots,i_{\pi(n)})$.\\
Generates $r'_i$ for $i=1,\ldots,n$ uniformly at random.\\
Sends $(E_{\textsf B}[{\mathbf B}_{i_{\pi(j)}} + r_{i_{\pi(j)}} + r'_{i_{\pi(j)}} \vert 4], E_{\textsf A}[r_{i_{\pi(j)}} + r'_{i_{\pi(j)}} \vert 4])$ for $j=1,\ldots,n$.\\
\STATE {\bf Party $\textsf A$:} \\
Decrypts $r_{i_{\pi(j)}} + r'_{i_{\pi(j)}}$ for $j=1,\ldots,n$.\\
Obtains $E_{\textsf B}[{\mathbf B}_{i_{\pi(j)}}\vert 5]$ $j=1,\ldots,n$.\\
Obtains $E_{\textsf B}[{\mathbf B}_{i_j}\vert 5]$ $j=1,\ldots,n$ through permutation by $\pi^{-1}$.\\
\STATE {\bf Party $\textsf A$:} \\
Simulates CWC for resulting $E_{\textsf B}[{\mathcal F}]$.
\end{algorithmic}
\end{algorithm}

\begin{figure*}[t]
\begin{center}
\includegraphics[width=1.2\textwidth]{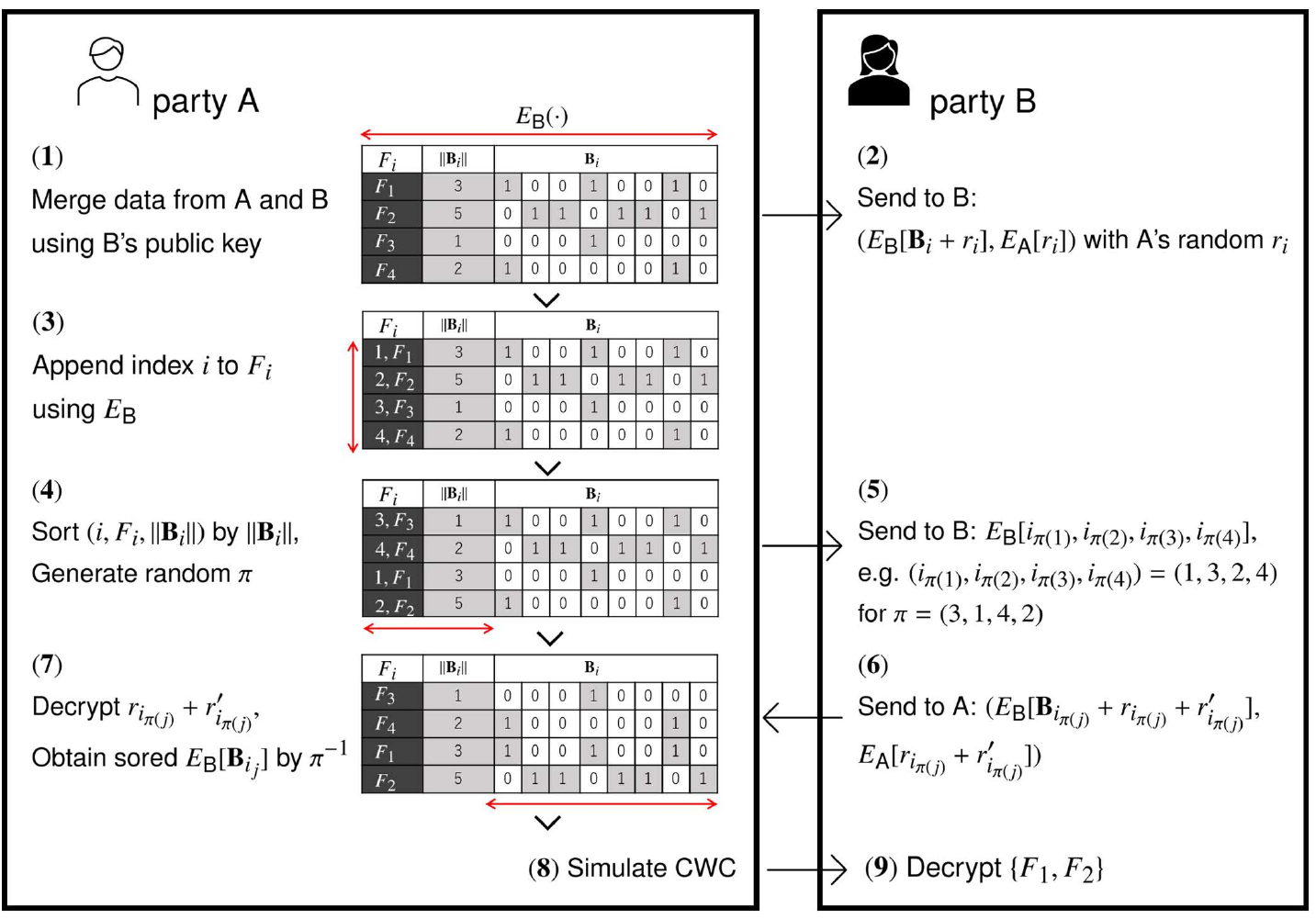}
\end{center}
\vspace{-1cm}
\caption{
An example run of Algorithm~\ref{new_sort}.
For simplicity, we omit the clock time in each ciphertext.
(1): Parties $\textsf A$ and $\textsf B$ jointly compute $\mathbf{B}_i$ and $\Vert\mathbf{B}_i\Vert$ for each feature $F_i$ (same as the baseline algorithm).
(2): $\textsf A$ securely sends $\mathbf{B}_i$; $\textsf B$ cannot learn anything.
(3): $\textsf A$ appends encrypted index $i$ for each $F_i$.
(4): $\textsf A$ sorts only $(F_i, \Vert\mathbf{B}_i\Vert)$ by $\Vert\mathbf{B}_i\Vert$.
(5): $\textsf A$ sends the sorted indices with random permutation; $\textsf B$ cannot learn anything.
(6): $\textsf B$ sends $\mathbf{B}_i$; $\textsf A$ cannot learn anything from it.
(7): $\textsf A$ decrypts the noise and obtain the correct order of ${\mathbf B}_{i_j}$; $\textsf A$ cannot learn anything.
(8): $\textsf A$ simulates CWC same as the baseline.
(9): Party $\textsf A,\textsf B$ share the resulting features.
}
\label{sort}
\end{figure*}

Sorting is a major bottleneck for private CWC. 
The reason for this is that pointers cannot be moved across ciphertexts.
For example, consider the case of secure integer sort. 
Let the variables $x$ and $y$ contain integers $a$ and $b$, respectively. 
In this case, by performing the secure operation $a<?b$, the result is obtained as $a<?b = c\in\{0,1\}$. 
Using this logical bit $c$, we can swap the values of $x$ and $y$ in $O(1)$ time
satisfying $x<y$ by the secure operation $x\leftarrow c\cdot a + \bar{c}\cdot b$ and
$y\leftarrow \bar{c}\cdot a + c\cdot b$.

In the case of CWC; however, each integer $i$ of feature $F_i$ is associated with the bit string $\mathbf{B}_i$.
Since any $x$ cannot be decrypted, we cannot swap the pointers appropriately. 
Therefore, the baseline algorithm swaps $\mathbf{B}_i$ explicitly.
As a result, the computation time for sorting increases to $O(mnk\log^2 k)$. 
Our main contribution of this study is to improve this complexity to $O(mnk+k\log^2 k)$
by reducing the cost for such explicit sorting.

Based on the FHE, we propose the improved secure CWC (Algorithm~\ref{new_sort}), 
which reduces the time complexity to $O(mnk+k\log^2 k)$.
An example run of Algorithm~\ref{new_sort} is illustrated in Fig.~\ref{sort}.
As shown in this example, the party $\textsf A$ can securely sort $k$ randomized features in $O(k\log^2 k)$ time
using a suitable {\em sorting network}, and then, according to the result of sorting,
$\textsf A$ swaps each associated bit string of length $nm$ in $O(kmn)$ time.
Following this preprocessing, the parties securely obtain minimal consistent features 
by decrypting the output of CWC.
Finally, we get the following result.

\begin{theorem}\label{complexity-proof}
Algorithm~\ref{new_sort} can simulate CWC in $O(kmn + k\log^2 k + k\log k\log mn)$ time and $O(kmn)$ space
under the assumption that FHE executes each bit operation in $O(1)$ time.
\end{theorem}
\begin{proof}
Compared to the baseline, the additional space is required for $\pi$ and $r_i$ and $r'_i$.
Thus, the space complexity remains $O(kmn)$.
For the time complexity, the main task is to sort $k$-triple $(F_i,\Vert\mathbf{B}_i\Vert,\mathbf{B}_i)$ 
in the increasing order of $\Vert\mathbf{B}_i\Vert$.
The improved algorithm sorts only the pairs $x_i = (F_i,\Vert\mathbf{B}_i\Vert)$ of integers,
where the size of $x_i$ is $O(\log k+\log mn)$ bits.
For each $x_i,x_j$, we can check if $\Vert\mathbf{B}_i\Vert \leq \Vert\mathbf{B}_j\Vert$ in $O(\log mn)$ time
and we can swap them in $O(\log k + \log mn)$ time using homomorphic operations in FHE.
It follows that the time for sorting all $x_i$ $(i=1,\ldots,k)$ is $O(k\log k (\log k + \log mn))$ time.
After sorting the pairs, the algorithm moves all $\mathbf{B}_i$ to the correct positions according to the rank of $x_i$ $(i=1,\ldots,k)$.
This cost is $O(kmn)$.
Therefore, time complexity is $O(kmn + k\log^2 k + k\log k\log mn)$.
\end{proof}

\begin{theorem}\label{security-proof}
Algorithm~\ref{new_sort} is secure under the assumption that the employed FHE is IND-CPA secure.
\end{theorem}
\begin{proof}
We show the security by constructing simulators for parties $\textsf A$ and $\textsf B$, respectively.

$\textsf B$'s view (what $\textsf B$ can obtain from $\textsf A$) is the following:
\begin{itemize}
\item $(E_{\textsf B}[{\mathbf B}_i + r_i \vert 1], E_{\textsf A}[r_i\vert 1])$ for $i=1,\ldots,n$;
\item $E_{\textsf B}[i_{\pi(1)}\vert 3], \ldots, E_{\textsf B}[i_{\pi(n)}\vert 3]$.
\end{itemize}
Their probability distributions are uniform and independent of each other.
Hence, the simulator for $\textsf B$ can replace them with
\begin{itemize}
\item $(E_{\textsf B}[\mathbf B_i + r''_i \vert 6], E_{\textsf A}[r''_i\vert 6])$ for $i=1,\ldots,n$ and $r''_i$, which are selected uniformly at random;
\item $E_{\textsf B}[\pi'(1)\vert 6], \ldots, E_{\textsf B}[\pi'(n)\vert 6]$ for $\pi' \in \mathfrak S_n$, which is selected uniformly at random.
\end{itemize}
Note that,
even if an adversary knows $\mathbf B_i$,
it is computationally impossible to distinguish between 
$E_{\textsf B}[\mathbf B_i + r''_i \vert 6]$ and $E_{\textsf B}[r''_i \vert 6]$
by the IND-CPA security of the cryptosystem $E_{\textsf B}$.

Next, we construct a simulator \texttt{Sim} for the party $\textsf A$.
Although what $\textsf A$ can obtain from $\textsf B$ is 
\[
  \left\{
    \left.
\left(
E_{\textsf B}[{\mathbf B}_{i_{\pi(j)}}+r_{i_{\pi(j)}} + r'_{i_{\pi(j)}} \vert 4],
E_{\textsf A}[r_{i_{\pi(j)}} + r'_{i_{\pi(j)}} \vert 4] 
\right)
\right\vert  j=1,\ldots,n
\right\}
\]
this is equivalent to $\{E_{\textsf B}[{\mathbf B_{i_j}}\vert 5] \mid j=1,\ldots, n\}$ after decryption and permutation.

On the other hand, the sequence $(i_1,\ldots,i_n)$ is not explicitly given to $\textsf A$,
and $\textsf A$ recognizes it through the alignment between
\begin{itemize}
\item $(E_{\textsf B}[\Vert{\mathbf B}_{i_1}\Vert \vert 3], \ldots, E_{\textsf B}[\Vert{\mathbf B}_{i_n}\Vert \vert 3])$ and
\item $(E_{\textsf B}[{\mathbf B}_{i_1} \vert 5], \ldots, E_{\textsf B}[{\mathbf B}_{i_n} \vert 5])$.
\end{itemize}

Therefore, we define $\textsf A$'s view to be
\[
  {\tt View}_{\textsf A} = \left\{ \left.
      \left(E_{\textsf B}[\Vert{\mathbf B}_{i_j}\Vert \vert 3], E_{\textsf B}[{\mathbf B}_{i_j}) \vert 5]\right)
        \right\vert j=1,\ldots,n \right\}
\]
with $\Vert{\mathbf B}_{i_1}\Vert \leq \cdots \leq \Vert{\mathbf B}_{i_n}\Vert$.

On the other hand, we define the view that ${\tt Sim}$ should generate as follows:
While $\textsf A$ can generate $\{E_{\textsf B}[\Vert{\mathbf B}_{i_j} \Vert \vert 3] \mid j=1,\ldots,n\}$
with $\Vert{\mathbf B}_{i_1}\Vert \leq \cdots \leq \Vert{\mathbf B}_{i_2}\Vert$,
$\textsf A$ needs $\textsf B$'s cooperation to generate
$\{ E_{\textsf B}[{\mathbf B}_{i_j} \vert 5 ] \mid j=1,\ldots,n\}$.
Without $\textsf B$'s cooperation, 
${\tt Sim}$ selects $\pi' \in \mathfrak S_n$ uniformly at random, and generates its own view to be
\[
{\tt View}_{\tt Sim} = \left\{ \left.
\left(
E_{\textsf B}[\Vert{\mathbf B}_{i_j}\Vert \vert 3], E_{\textsf B}[{\mathbf B}_{\pi''(j)}\vert *]
\right)\right\vert
j=1,\ldots,n
\right\}.
\]
\texttt{Sim} can compute $E_{\textsf B}[{\mathbf B}_{\pi''(j)}\vert *]$ from $E_{\textsf B}[0\vert *]$ and
$E_{\textsf B}[{\mathbf B}_{\pi''(j)}\vert 0]$ taking advantage of the homomorphic property of the encryption system $E_{\textsf B}$.

Furthermore,
we define a distinguisher $\cal D$ as a PPT Turing machine which tries to distinguish between ${\tt View}_{\textsf A}$
and ${\tt View}_{\tt Sim}$ on input of 
$\{
( E_{\textsf B}[\Vert {\mathbf B}_i\Vert \vert 0], E_{\textsf B}[ {\mathbf B}_i \vert 0 ] ) \mid i=1,\ldots,n
\}.$

When we let $\Pr[Y = \textsf A \mid X= \textsf A] = 1/2 + \alpha_1$ and
$\Pr[Y = {\tt Sim} \mid X={\tt Sim}] = 1/2 + \alpha_2$,
the advantage of $\cal D$ is defined  as $\alpha_1 + \alpha_2$.

We show that , if $\cal D$'s advantage $\alpha$ is not negligible, we can construct a PPT attacker ${\tt Attck}$
that can brake the IND-CPA security of the encryption system $E_{\textsf B}$ with non-negligible advantage.
Our attacker ${\tt Attck}$ plays the IND-CPA game exploiting an oracle ${\cal O}_{\text{IND}}$ as follows:
\begin{enumerate}
\item ${\tt Attck}$ generates ${\mathbf B}_1, {\mathbf B}_2$ with $\Vert{\mathbf B}_1\Vert\leq \Vert{\mathbf B}_2\Vert$;
\item ${\tt Attck}$ lets $x_1= \Vert{\mathbf B}_1\Vert$ and
  $x_2=\Vert{\mathbf B}_2\Vert$ and throws a query $(x_1,x_2)$ to ${\cal O}_{\text{IND}}$;
\item ${\cal O}_{\text{IND}}$ selects $i\in\{1,2\}$ uniformly at random and sends $c=E_{\textsf B}[x_i \vert -1]$ to ${\tt Attck}$;
\item ${\tt Attck}$ initializes $\cal D$ by inputting 
$(E_{\textsf B}[\Vert{\mathbf B}_1\Vert \vert 0], E_{\textsf B}[\Vert{\mathbf B}_1\Vert \vert 0]),
(E_{\textsf B}[\Vert{\mathbf B}_2\Vert \vert 0], E_{\textsf B}[\Vert{\mathbf B}_2\Vert \vert 0])$;
\item {\bf First query.} ${\tt Attck}$ throws to $\cal D$ the query: 
$(E_{\textsf B}[\Vert{\mathbf B}_2\Vert \vert 2], c),
(E_{\textsf B}[\Vert{\mathbf B}_2\Vert \vert 2], E_{\textsf B}[{\mathbf B}_2 \vert 2])$;
\item If $\cal D$ replies with $\textsf A$, ${\tt Attck}$ outputs $1$ and terminates.
\item {\bf Second query.} ${\tt Attck}$ generates $c'$ by adding $E_{\textsf B}[0\vert 3]$ to $c$.
Note that ${\cal D}_{\textsf B}(c') = {\cal D}_{\textsf B}(c)$ holds.
${\tt Attck}$ throws to $\cal D$ the query:
$(E_{\textsf B}[\Vert{\mathbf B}_1\Vert \vert 3],E_{\textsf B}[{\mathbf B}_2 \vert 3]), (E_{\textsf B}[\Vert{\mathbf B}_2\Vert \vert 3], c')$.
\item If $\cal D$ replies with ${\tt Sim}$, ${\tt Attck}$ outputs $1$ and terminates.
\item ${\tt Attck}$ outputs $2$.
\end{enumerate}

We evaluate ${\tt Attck}$'s advantage as follows.
We assume ${\cal D}_{\textsf B}(c)=x_1$.
The probability of this case is $1/2$.
The probability that $\cal D$ replies with $\textsf A$ to the first query or $\cal D$ replies with ${\tt Sim}$ to the second query is
\[
\frac{1}{2} + \alpha_1 + \frac{1}{2} + \alpha_2 - (\frac{1}{2}+\alpha_1)(\frac{1}{2}+\alpha_2) = \frac{3}{4} + \frac{\alpha}{2}-\alpha_1\alpha_2
\geq \frac{3}{4} + \frac{\alpha}{4},
\]
since the first and second queries are mutually independent.

When assuming $D_B(c)=x_2$, we see that $\Pr[\text{$\cal D$ outputs ${\tt Sim}$ at the first query }] - 1/2$ is negligible.
Otherwise, $\cal D$ can be used as an attacker to break the IND-CPA security of $E_B$.  
Therefore, $\Pr[{\text{\texttt{Attck} outputs 2}}] - 1/4$ is negligible.
Consequently, we have
\[
\Pr[\text{${\tt Attck}$'s guess is right}]\geq 
\frac{1}{2}\left( \frac{3}{4} + \frac{\alpha}{4} \right) + \frac{1}{2}\cdot \frac{1}{4} = \frac{1}{2} + \frac{\alpha}{8}.
\]
Since we assume that $\alpha$ is not negligible, neither is $\alpha/4$.
\end{proof}

\section{Experiments}
We implemented the baseline and improved algorithms for secure CWC in \texttt{C++} using TFHE library\footnote{\tt https://tfhe.github.io/tfhe}.
The experiments were carried out on a machine equipped with Intel Core i7-6567U (3.30GHz) processor and 16GB of RAM.
In the following, $m$ (resp. $n$) is the number of positive (resp.\ negative) data and $k$ is the number of features.

Table~\ref{tab:cwc_section_total_time} summarizes the running time of the baseline algorithm (naive implementation of Algorithm~\ref{algo1}
using TFHE) for random data
generated for $k\in\{10, 50,100\}$ and $mn\in\{100,500,1000\}$.
The complexity analysis shows that the running time increases in proportion to $mn$. 
This experimental result confirms this in real data.
The table clearly shows that the sorting process is the bottleneck.

\begin{table}[t]
  \begin{center}
  \caption{Running time (sec) of baseline algorithm (naive secure CWC). Task 1: computing $\mathbf{B}_i$'s. Task 2: sorting $\mathbf{B}_i$'s. Task 3: feature selection.}
  	\label{tab:cwc_section_total_time}
		\begin{tabular}[t]{rrrrrr} \hline
		\qquad $k$ & \qquad $mn$ & \qquad\qquad Task~1 & \qquad\qquad Task~2 & \qquad\qquad Task~3 & \;\;\;\;  \\ \hline
		 & 100 & 60.3 & {\bf 835.8} & 111.9 \\ 
		 10 & 500 & 300.5 & {\bf 4,252.4} & 558.1 \\ 
		 & 1,000 & 601.4 & {\bf 8,867.0} & 1,114.2 \\ \hline 
		 & 100 & 301.8 & {\bf 6,292.6} & 589.3 \\ 
		 50 & 500 & 1,502.9 & {\bf 30,364.6} & 2,941.0 \\ 
		 & 1,000 & 3,007.0 & {\bf 62,124.6} & 5,919.7 \\ \hline 
		 & 100 & 603.7 & {\bf 16,148.5} & 1,179.0 \\ 
		 100 & 500 & 3,005.9 & {\bf 76,315.2} & 5,952.5 \\ 
		 & 1,000 & 6,014.1 & {\bf 154,143.5} & 11,867.0 \\ \hline
	\end{tabular}
  \end{center}
\end{table}

Table~\ref{tab:naive-vs-improve} compares the running time of preprocessing in baseline and improved algorithms.
According to the results, the proposed algorithm significantly improves the bottleneck in naive CWC for secure computing.
We should note that baseline and improved algorithms both compute exactly the same solution as the CWC on plaintexts.
We also show the details of improved algorithm:
`sorting' means the time for sorting of the triples $(F_i,||{\mathbf B}_i||,i)$ of integers.
`other task' means the time for remaining tasks including generating/adding/subtracting random noise $r_i$, moving ${\mathbf B}_i$, decrypting integers, etc.

\begin{table}[t]
  \begin{center}
  \caption{Running time (sec) of baseline and improved algorithms.
  `baseline' is same as Task 2 in Table~\ref{tab:cwc_section_total_time} (i.e. the bottleneck).
  `improved:' is the running time of corresponding task in the improved algorithm, where
  `sorting' and `other tasks' are the details.
  }
  	\label{tab:naive-vs-improve}
\scalebox{0.9}[0.9]{ 
		\begin{tabular}[t]{rrrrrr}\hline
		\qquad $k$ & \qquad $mn$ & \qquad {\bf baseline} & \qquad {\bf improved:} & \qquad sorting & \qquad other tasks  \\ \hline 
		 & 100 & {\bf 835.8}  & {\bf 203.7}	& 69.4 & 134.2 \\
		 10 & 500 & {\bf 4,252.4}	& {\bf 286.2}	& 89.5	& 196.6 \\
		 & 1000 & {\bf 8,867.0}	& {\bf 302.5}	& 98.9	& 203.6 \\ \hline
		 & 100 & {\bf 16,148.5}	& {\bf 3,311.2}	& 1,865.5	& 1,445.7 \\
		 100 & 500 & {\bf 76,315.2}	& {\bf 4,601.9}	& 2,647.7	& 1,954.1 \\
		 & 1000 & {\bf 154,143.5}	& {\bf 4,671.4}	& 2,660.8	& 2,010.5 \\ \hline
	\end{tabular}
}
  \end{center}
\end{table}

Table~\ref{tab:UCI} displays the running time of improved algorithm for real data available from UCI Machine Learning Repository\footnote{\tt https://archive.ics.uci.edu/ml/index.php},
because since these datasets contain more than three feature/class values, we treated them as a binary classification between one feature/class 
and the other.

\begin{table}[t]
  \begin{center}
  \caption{Running time (sec) of improved algorithm for real data in UCI Machine Learning Repository.
  }
  	\label{tab:UCI}
		\begin{tabular}[t]{rrrrrrr}\hline
		\qquad dataset & \qquad $k$ & \quad $mn$ & \qquad time & \qquad sorting & \qquad other tasks \\ \hline
		 {\tt Letter} & 16 & 196  & {\bf 252.2} &  80.6 & 171.4 \\
		 {\tt Breast Cancer} & 10 & 2,464 	&	{\bf 312.6} & 103.1 & 209.5 \\
		 {\tt Covertype} &  54 & 	979 &  {\bf 1,653.5} &  836.9	& 816.6 \\ \hline
	\end{tabular}
  \end{center}
\end{table}

We demonstrated that the proposed algorithm works well for real-world multi-level feature selection problems.
We only evaluated the running time in this experiment, but the relevance of the extracted features 
is guaranteed because the secure CWC algorithm produces the same solution as the original~\cite{Shin2009}.

\section{Conclusion}

On the basis of fully homomorphic encryption, 
we proposed a faster private feature selection algorithm
that allow us to securely compute functional features from distribute private datasets.
Our algorithm can simulate the original CWC algorithm, which chooses favorable features by sorting.
In addition to the improvement in computational complexity, the proposed algorithm solves 
the private feature selection problem in practical time for a variety of real data.
One of the remaining challenges is to improve sorting at a lower cost 
because CWC does not always require exact sorting.
Then, ambiguous sorting possibly reduces the computation time maintaining solution quality.
At this time, the proposed algorithm is not applicable to real number for feature value.
This is because TFHE is not good at floating point operations. 
Extending the TFHE library to enable secure feature selection for real-valued data is a future challenge.



\end{document}